\documentclass[english]{article}
\usepackage[T1]{fontenc}
\usepackage[latin9]{inputenc}
\usepackage{amsmath}
\usepackage{amssymb}

\makeatletter

\makeatletter

\newtheorem{theorem}{Theorem}[section]
\newtheorem{lemma}{Lemma}[section]

\date{}

\makeatother

\begin{document}

\title{Processes with a Local Deterministic Interaction: Invariant Bernoulli
Measures}

\author{V.A.~Malyshev, A.A.~Zamyatin }
\maketitle
\begin{abstract}
A general class of Markov processes with a local interaction is introduced,
which includes exclusion and Kawasaki processes as a very particular
case. Bernoulli invariant measures are found for this class of processes. 
\end{abstract}


\section{Introduction}

The goal of this paper is to introduce a class of processes with a
local interaction, which we call exchange or Boltzmann processes as
they model transformations of the internal degrees of freedom and/or
chemical reactions for pairs of particles. The introduced class of
processes includes such well-known processes as exclusion processes
and Kawasaki processes.

The definition is as follows. Consider a graph $G$, finite or countable,
with the set of vertices $V=V(G)$ and the set of edges $L=L(G)$.
We define configuration as a function $x_{v},\, v\in V,$ on the set
of vertices with values in some set $X$. The set $X$ can be interpreted
as the set of all characteristics of a site $v\in V(G)$ and/or of
the particles sitting at $v$ (such as the types of particles, their
form, energy etc.).

On the set $X^{V}$ of configurations a continuous time Markov process
$\xi_{t}=\xi_{t}^{G,F,\{\lambda_{l}\}}$ is defined as follows. The
transitions occur for each edge $l=(v,v')\in L(G)$ with rate $\lambda_{l}=\lambda_{l}(x_{v},x_{v'})$,
independently of all other edges. For a given~$l$, the transition
is a simultaneous transformation (binary reaction) of the spins $x_{v}$
and $x_{v'}$, \begin{equation}
(x_{v}(t),x_{v'}(t))\to(x_{v}(t+dt),x_{v'}(t+dt))=F(x_{v}(t),x_{v'}(t))\label{Fmap}\end{equation}
 where $F:S=X\times X\to S=X\times X$ is some fixed mapping. We always
assume $F$ to be symmetric, that is $Fj=jF$, where $j(x_{1},x_{2})=(x_{2},x_{1})$.
This explains why the order of vertices in (\ref{Fmap}) does not
play any role. Thus, the process on $G$ is defined by a function
$F$ and by the set of functions $\lambda_{l}(x_{v},x_{v'})=\lambda_{l}(x_{v'},x_{v})$,
also assumed to be symmetric.

If the set $X$ and the graph $G$ are finite, then this defines a
finite continuous time Markov chain, which we denote by $\xi_{t}$.
Otherwise, for the existence of the process, one should impose some
weak restrictions on $F$ and $\lambda_{l}$.

The introduced process is a process with local interaction, these
processes play nowadays an important role in constructing physical
models, see for example \cite{Kipnis,Liggett,Spohn}. A particular
case are Kawasaki processes, where a pair of points exchanges spins,
that is $F$ is a permutation. Even more popular are exclusion processes,
where $X=\{0,1\}$ and $F$ is also a permutation. In general the
choice of $F$ should correspond to the transformation of degrees
of freedom of neighbour particles (for example, of water molecules)
or to chemical reactions. As far as we know, such processes were never
studied in sufficient generality.

The first problem we are solving here: for given $F$ and $\lambda_{l}$,
describe all invariant Bernoulli measures (IBM). A measure $\mu$
on $X^{V}$ is called Bernoulli, if for some probability measure $\nu$
on $X$ we have \[
\mu=\nu\times\nu\times\ldots=\nu^{V}.\]
 Such measures are well-known and are very important for the study
of exclusion processes, see \cite{Liggett}.

\section{Invariance criteria}

We assume here that $X$ and $G$ are finite, and $F$ is assumed
to be one-to-one. Due to compactness, at least one invariant measure
always exists. Moreover, if $F$ is one-to-one, then the uniform measure
on $X^{V}$ is invariant. We want to know for which maps $F$ there
exist other invariant Bernoulli measures.

Let $\nu$ be a probability measure on $X$. Let us consider the measure
$\nu^{2}=\nu\times\nu$ on the set $S=X\times X.$ Let $\nu^{2}$
take $k$ different values, $d_{1},d_{2},\ldots,d_{k}$. Define a
partition $\{S_{i}\}$ of the set $S$ such that $S_{i}$ consists
of all points of $S$ where $\nu^{2}$ takes the value $d_{i}$.

Note that the map $F$ can be uniquely expanded on a finite number
of cycles on $S$. Let $C_{1},\ldots,C_{n}$ be the supports of these
cycles. 
We say that the measure $\nu$ agrees with the map $F$ if the support
of any cycle of $F$ belongs to only one of the sets $S_{i},$ $i=1,\ldots,k.$
In other words, the partition $\{C_{j}\}$ is finer than the partition
$\{S_{i}\}$: for any $C_{j}$ there exists $S_{i}$ such that $C_{j}\subseteqq S_{i}$.

The following result gives a convenient criterion to check whether
a given Bernoulli measure is invariant for given $F$.

\begin{theorem}\label{1} Let finite $X$, $G$ be given and let
$F$ be one-to-one. Assume that for any edge $l$ the rates $\lambda_{l}=\lambda_{l}(s)$
satisfy the condition: $\lambda_{l}(s)=\lambda_{l}(F^{-1}(s))$ for
all $s\in S.$ Then the following conditions are equivalent:
\begin{itemize}
\item [{\rm 1.}] Bernoulli measure $\mu=\nu\times\nu\times\ldots=\nu^{V}$
is an invariant measure of the Markov process $\xi_{t}$ for any finite
graph $G$. 
\item [{\rm 2.}] The measure $\nu^{2}$ is an invariant measure of
the Markov process $\xi_{t}$ for the graph $G_{2}$ with two vertices
and one edge between them. 
\item [{\rm 3.}] The measure $\nu^{2}$ is invariant with respect to
$F$. 
\item [{\rm 4.}] The measure $\nu$ agrees with $F$ in the sense defined
above. 
\end{itemize}
\end{theorem}

\textbf{Proof} \ Firstly, it is evident that conditions 3 and 4 are
equivalent, that is, the invariance of $\nu^{2}$ with respect to
$F$ is equivalent to the fact that $\nu^{2}$ takes constant values
on the support of each cycle of $F$. It is also evident that $1\Rightarrow2$.

Let us prove then that $2\Rightarrow1$. 
In fact, let $G,X,F$ be given and let $l$ be the edge with vertices
$v$ and $v'$. Denote $\xi_{t}^{(l)}$ the Markov chain on $X^{V}$
in which the only transitions are at the edge $(v,v')$: \[
(x_{v},x_{v'})\to F(x_{v},x_{v'}),\]
 with rates $\lambda_{l}(x_{v},x_{v'})$, that is, the remaining $\lambda_{l'}(x_{v},x_{v'})=0,\, l'\neq l$.
It is clear that if the condition 2 holds, then the Bernoulli measure
$\nu^{V}$ on $G$ is invariant with respect to any Markov chain $\xi_{t}^{(l)}$.
Then we get the assertion from the following general and evident proposition.
Let a collection $\mathbf{\xi}_{t}^{(l)}$ of Markov processes on
the same state space $A$ be given, with the rates $\lambda_{\alpha\beta}^{(l)},\,\alpha,\beta\in A,\,\alpha\neq\beta,$
correspondingly. Let moreover all the processes have the same invariant
measure $\pi=\{\pi_{\alpha}\}$. Then the Markov process on $A$ with
rates $\mu_{\alpha\beta}=\sum_{l}\lambda_{\alpha\beta}^{(l)},\,\alpha\neq\beta,$
has the same invariant measure. The proof of this proposition is immediately
obtained by summing over $l$ the equations for stationary probabilities
of $\mathbf{\xi}_{t}^{(l)}$.

Let us prove now that $3\Rightarrow2$, that is, if $\nu^{2}(s)=\nu^{2}(F^{-1}(s))$
for any $s\in S$ then the measure $\nu^{2}$ is an invariant measure
of the Markov process $\xi_{t}$ for the graph $G_{2}$ with two vertices
and one edge between them. To do this, let us write down the equations
for the stationary probabilities of the Markov chain on $G_{2}$:
\[
\lambda_{l}(s)\nu^{2}(s)=\lambda_{l}(F^{-1}(s))\nu^{2}(F^{-1}(s)),\quad s\in S.\]
 They evidently hold under our assumptions. This implies $2\Rightarrow3$
as well. 
 $\square$

\section{Description of invariant measures}

Theorem~\ref{1} allows for given $F$ to check whether a given Bernoulli
measure~$\nu^{V}$ is invariant or not. To classify all invariant
Bernoulli measures for given $F$ is a more complicated problem. We
give now simple combinatorial algorithms which allow, for given $F$,
to construct all IBM.

Let a measure $\nu$ take values $a_{1},\ldots,a_{m}$, all different.
Denote $X_{i}=\{x:\nu(x)=a_{i}\}$ and $S_{ij}=(X_{i}\times X_{j})\cup(X_{j}\times X_{i})=S_{ji}$.
We say that a measure $\nu$ is a general situation measure if all
pairwise products $a_{i}a_{j}$ are different. The partition $X_{1},\ldots,X_{m}$
defines a $(m-1)$-parametric family $\{(a_{1},\ldots,a_{m}),$ $a_{1}+\ldots+a_{m}=1\}$
of general situation measures. We say that an IBM $\nu^{V}$ is a
general situation measure if $\nu$ is a general situation measure.
We will describe all such measures, under some assumptions. Let us
note that for general situation measures any cycle $C_{k}$ of $F$
belongs to only one set $S_{ij}$, that is, all $S_{ij}$ are different.

A set $A\subset S$ is called connected, if for any two elements $(a,b),(a',b')\in A$
there exists a chain of elements $(a_{i},b_{i})\in A$, \[
(a,b)=(a_{1},b_{1}),\,(a_{2},b_{2}),\,(a_{3},b_{3}),\ldots,(a_{n},b_{n})=(a',b')\]
 in which all subsequent pairs have a common element, that is, \[
(\{a_{i}\}\cup\{b_{i}\})\cap(\{a_{i+1}\}\cup\{b_{i+1}\})\neq\emptyset,\quad i=1,\ldots,n-1.\]
 Obviously, any set $B$ can be uniquely partitioned into connected
components.

We shall give an algorithm for constructing all general situation
IBM in case when all the cycles of $F$ are connected.

\begin{theorem}\label{2} If all the cycles of $F$ are connected
then, among the partitions agreeing with $F$, there exists a unique
minimal partition $\{S_{ij}=(X_{i}\times X_{j})\cup(X_{j}\times X_{i})\}$.
Any general situation IBM belongs to the family of IBM defined by
this minimal partition. The minimal partition $\{S_{ij}\}$ is constructed
by the algorithm given in the proof. 
\end{theorem}

\textbf{Proof} \ A set $A\subset S$ is called half-admissible, if
it can be represented as $A=X_{1}\times X_{2}$, where either $X_{1}=X_{2}$
or $X_{1}\cap X_{2}=\emptyset$. A set $A\subset S$ is called admissible,
if it can be represented as $A=(X_{1}\times X_{2})\cup(X_{2}\times X_{1})$,
where either $X_{1}=X_{2}$ or $X_{1}\cap X_{2}=\emptyset.$ Half-admissible
and admissible sets are always connected.

\begin{lemma} \label{lem} For any connected set $B\subset S$, among
admissible sets containing (covering) $B$ there exists a unique minimal
admissible set covering $B$. \end{lemma}

\textbf{Proof of Lemma \ref{lem}} The proof of this lemma consists
in direct construction of such covering set. Let us construct first
a half-admissible set $X_{1}\times X_{2}$ containing $B$. To do
this, take some element $(a,b)\in B$. Put, for example, $a\in X_{1},b\in X_{2}$.
If $a=b$, then it follows that $X_{1}=X_{2}$. In this case the minimal
set will be the set $X_{1}\times X_{1}$, where $X_{1}$ is the projection
of $B$ on $X$ (the projection of the set $B\subset S=X\times X$
on $X$ is the set of all elements $x\in X$ such that there exists
$a\in X$ such that either $(a,x)$ or $(x,a)$ belongs to $B$).

Consider now the case when $a\neq b$. Then necessarily $b\in X_{2}$,
and also for all $(a,x)\in B$ necessarily $x\in X_{2}$. Continuing
this process, due to connectedness of $B$, we encounter all elements
of $B$ and will construct $X_{1}$ and $X_{2}$. During this process
it can occur that some element $c$ belongs both to $X_{1}$ and to
$X_{2}$. Then, by definition of half-admissible set, it should be
$X_{1}=X_{2}$. The symmetrized set $\overline{B}=(X_{1}\times X_{2})\cup(X_{2}\times X_{1})$
is called the closure of the set $B$. The lemma is proved. $\square$

We return to the proof of Theorem \ref{2}. Let $\overline{C_{i}}$
be the closure of the cycle $C_{i}$. To each cycle $C_{i}$ there
corresponds a symmetric cycle $C_{i}^{sym}$, where all elements of
$C_{i}^{sym}$ are the permutations of the elements of $C_{i}$. Moreover,
either $C_{i}^{sym}=C_{i}$ or $C_{i}^{sym}\cap C_{i}=\emptyset$.
Then $D_{i}=\overline{C_{i}^{sym}\cup C_{i}}$ define a covering of
the set $S$, however, they can intersect with each other. If some
$D_{1}$ and $D_{2}$ intersect, then their union is connected. In
this case one can take $\overline{D_{1}\cup D_{2}},D_{3},\ldots,D_{m}$
instead of the collection of sets $D_{1},D_{2},\ldots,D_{m}$. On
each step of this procedure the number of sets in the covering diminishes
by $1$, and finally we get a system of non-intersecting admissible
sets which defines the partition $\{X_{i}\}$, and thus all general
situation IBM. The resulting partition does not depend on the order
in which we choose the pairs of intersecting subsets, since at each
step we take the minimal admissible set. The theorem is proved. $\square$

If there exist non-connected cycles, then several families of IBM
are possible, as Example~\ref{e3} below shows. An algorithm for
constructing all such families is similar, but more involved, we discuss
it below.

\textbf{Example 1.} \ \label{e1} 
Let us consider Kawasaki processes, when $F$ is the permutation.
In this case all the cycles have the length 1 or 2. Then each $X_{i}$
consists of one point only, and each $S_{ij}$ consists of one (if
$i=j$) or two (if $i\neq j$) elements. Then any Bernoulli measure
is invariant.

\textbf{Example 2.} \ \label{e2} Let $F(a,b)=(f(a),f(b))$, 
where $f$ is a one-to-one mapping $X\to X$ having the cycles $X_{i}$.
Then $S_{ij}=(X_{i}\times X_{j})\cup(X_{j}\times X_{i})$ define all
IBM.

\textbf{Example 3.} \ \label{e3} The simplest example when there
exist two general situation IBM is as follows. Let $X$ consist of
four points, that is $X=\{x_{1},x_{2},x_{3},x_{4}\}$. Let any cycle
of $F$ consist of one point only, except for the following two cycles:
\[
C_{1}=\{(x_{1},x_{2}),(x_{3},x_{4})\}\subset S\]
 and the symmetric one, \[
C_{1}^{symm}=\{(x_{2},x_{1}),(x_{4},x_{3})\},\]
 consisting of two points. Then there are two admissible partitions,
$X_{1}=\{x_{1},x_{3}\},X_{2}=\{x_{2},x_{4}\}$ and $X'_{1}=\{x_{1},x_{4}\},X'_{2}=\{x_{2},x_{3}\}$,
which define two one-parametric families of general situation IBM.

Further on, we shortly describe the algorithm for constructing all
families of IBM in the general case. Let us show first that for any
set $B\subset S$ there exists a unique minimal covering (that is,
the covering belonging to any such covering of the set $B$) by non-intersecting
half-admissible sets. In fact, if $B$ is not connected, then for
any its connected component $B_{i}$ consider the closure $\overline{B_{i}}$.
It is easy to see that $\overline{B_{i}}$ do not intersect. Then
the minimal half-admissible sets covering $\overline{B_{i}}$ do not
intersect as well.

Let $C_{i}$ be all the cycles of $F$, and $C_{ij}$ be all connected
components of the cycle $C_{i}$. Take the closure $\overline{C_{ij}}$
of each $C_{ij}$. Firstly, for any $i$ we construct a minimal admissible
set $A_{i}$ containing all $\overline{C_{ij}}$. The problem is that
there can be several such $A_{i}$ (see Example~\ref{e3}). Then
for given $A_{i}$ we construct, as above, (already unique) minimal
partition $\{S_{kl}\}$ such that each $A_{i}$ belongs to one of
the sets $S_{kl}$.

\section{Generalizations and remarks}

\noindent \textsl{\underline{Maps which are not one-to-one}}

\smallskip{}

Let us consider the case when $F$ is not one-to-one. A point $s\in S$
is called cyclic if $s=F^{n}(s)$ for some $n>0$, where $F^{n}$
is the $n$th iteration of the map~$F$. The set of cyclic points
is subdivided onto cycles. The remaining points are called inessential.

From the definition of invariant (with respect to $F$) measure it
easily follows that the invariant measure is zero on the set of inessential
points. Let a measure $\nu$ on $X$ take values $a_{0}=0,\, a_{1},\ldots,a_{k}$.
Put $X_{i}=\{x\in X:\nu(x)=a_{i}\},\, X_{0}\neq\emptyset$. Then $(X_{0}\times X)\cup(X\times X_{0})$
contains the set of inessential points and possibly also some cycles.
Let $C_{1},C_{2},\ldots,C_{m}$ be all the cycles of the map $F$
which do not belong to $(X_{0}\times X)\cup(X\times X_{0})$.

If, instead of $X$, we consider the set $X\setminus X_{0}$, then
$(X\setminus X_{0})\times(X\setminus X_{0})$ is invariant with respect
to $F$ and is the union of the cycles $C_{1},C_{2},\ldots,C_{m}$,
and, moreover, the map $F$ on $(X\setminus X_{0})\times(X\setminus X_{0})$
is one-to-one. It means that the description of invariant measures
can be reduced to the case when $F$ is one-to-one.

\medskip{}

\noindent \textsl{\underline{Countable $X\vphantom{p}$}}

\smallskip{}

This case is quite similar to the case when $X$ is finite. In fact,
if $F$ has infinite cycles, then $\nu^{2}$ should be zero on them.
Thus, one can delete from $X$ the projections of all infinite cycles,
that is we can restrict ourselves to the case when all cycles are
finite. All the rest is similar to the case of finite $X$.

\medskip{}

\noindent \textsl{\underline{Oriented graph}} \nopagebreak\smallskip{}

Our results take place also in a more general case when the graph
$G$ is oriented. The map $F$ from $S=X\times X$ to itself is not
necessary symmetric. Moreover the intensities $\lambda_{l}(x_{v},x_{v'})$
may be non-symmetric functions of the spin values. However, the condition
$\lambda_{l}(x_{v},x_{v'})=\lambda_{l}(F^{-1}(x_{v},x_{v'}))$ should
be fulfilled.

\medskip{}
 \textsl{\underline{Links with physics}}

\noindent \smallskip{}

The introduced processes have many links with physics, on the intuitive
level. For example, the book \cite{Frenkel} explains many facts of
behaviour of liquids and amorphous bodies using stochastic exchange
interaction between nearby molecules. It gives an alternative to the
common approach based on hard-balls-type models.

However when one tries to derive an exchange process from the existing
fundamental physical theory, one encounters many difficulties. For
example, what is the set of states (that is, the set $X$ introduced
above) for the water molecules? The simplest classical model of water
molecule includes at least the lengths of segments $OH$ and the angles
between them. The known quantum mechanical model is even more complicated:
it defines a tetrahedron using molecular orbitals \cite{Franks}.
Moreover, it is known that chemical bonds of hydrogenic character
may appear between closely situated water molecules --- one molecule
can have up to 4 such bonds. They form a random graph of bonds (edges)
which presumably defines such properties of water as density, viscosity,
heat capacity etc., and their abnormal character comparing with other
liquids. In this context one can remark that the time dependence,
even random, $\lambda_{l}=\lambda_{l}(t)$ does not change the invariant
measures, under keeping the symmetry condition for any~$t$. This
means also that the graph $G$ itself (depending on the configuration
of particles in the space) does not play an important role, since
we can consider the complete graph where some $\lambda_{l}=0$.

\medskip{}

\noindent \textsl{\underline{Some problems}}

\smallskip{}

It could be interesting to get similar results for the case when $X$
is a smooth manifold. In this case conservation laws play an important
role. For finite $X,G$, an additive conservation law is a function
$E$ on $X$ such that if $F(x,y)=(x_{1},y_{1})$ then \[
E(x)+E(y)=E(x_{1})+E(y_{1}).\]
 If $\nu$ is an IBM and $C$ is an arbitrary constant, then the conservation
law is \[
E(x)=C\ln\nu(x).\]

Other interesting possibilities are when $F$ is random (see some
examples in \cite{MalPir}) or even when $F$ is quantum and $X$
is a Hilbert space.

\end{document}